\documentclass[12pt]{iopart}

\usepackage{iopams}  
\usepackage{cite}
\usepackage[]{graphicx}

\newcommand{\micron}{$\mu$m}

\newcommand{\kT}{k_{\rm B}T}

\newcommand{\la}{{\langle \hspace{0.1em} l \hspace{0.1em} \rangle}} 
\begin{document}

\title[]{Dynamics in Steady State In Vitro Acto-Myosin Networks}

\author{Adar Sonn-Segev,$^{1}$ Anne Bernheim-Groswasser,$^{2}$ Yael Roichman$^{1*}$}

\address{$^1$Raymond \& Beverly Sackler School of Chemistry, Tel Aviv University, Tel Aviv 6997801, Israel}
\address{$^2$Department of Chemical Engineering, Ilse Kats Institute	for Nanoscale Science and Technology, Ben Gurion University of the Negev, Beer-Sheva 84105, Israel}
\ead{roichman@tau.ac.il}
\vspace{10pt}
\begin{indented}
\item[]July 2014
\end{indented}

\begin{abstract}

It is well known that many biochemical processes in the cell such as gene regulation, growth signals and activation of ion channels, rely on mechanical
stimuli. However, the mechanism by which mechanical signals propagate through cells is not as well understood. In this review we focus on stress propagation in a minimal model for cell elasticity, actomyosin networks, which are comprised of a sub-family of cytoskeleton proteins. After giving an overview of th actomyosin network components, structure and evolution we review stress propagation in these materials as measured through the correlated motion of tracer beads. We also discuss the possibility to extract structural features of these networks from the same experiments. We show that stress transmission through these networks has two pathways, a quickly dissipative one through the bulk, and a long ranged weakly dissipative one through the pre-stressed actin network.    

\end{abstract}

%
\vspace{2pc}

\section{Structure and components of In vitro acto-mysoin networks}
Two of the central functions of the cytoskeleton are to support the cell's shape and to control its motion. These roles are done predominantly by variants of actomyosin networks consisting mainly of a structural protein actin, and a molecular motor, myosin. Additional actin binding proteins may modify the structure of actin networks and their polymerization kinetics.
 
\subsection{Network building blocks} 
\subsubsection*{Actin and actin polymerization} 
Actin is an abundant and highly conserved protein in most eukaryotic cells\cite{Dominguez2011}. Actin networks are responsible for many cellular processes such as cell motility, cell division, and the maintenance of cellular integrity\cite{Dominguez2011,Pollard2009,Cooper2003,Blanchoin2014}. To accommodate all of these roles actin networks adopt a variety of structures within cells ranging from finely branched networks in the lamellipodia to thick bundles in stress fibers\cite{Blanchoin2014}. It is becoming increasingly clear that cells control the structure and kinetics of their actin networks by numerous actin binding proteins (ABP). These ABP are involved in many processes including nucleation, elongation, and branching of actin filaments, in filament disassembly via severing and depolymerization, and in bundling and crosslinking of filaments into fibers and unorganized networks\cite{Dominguez2011,Blanchoin2014}. 

\begin{figure}[h]
	\centering
	\includegraphics[scale=0.5]{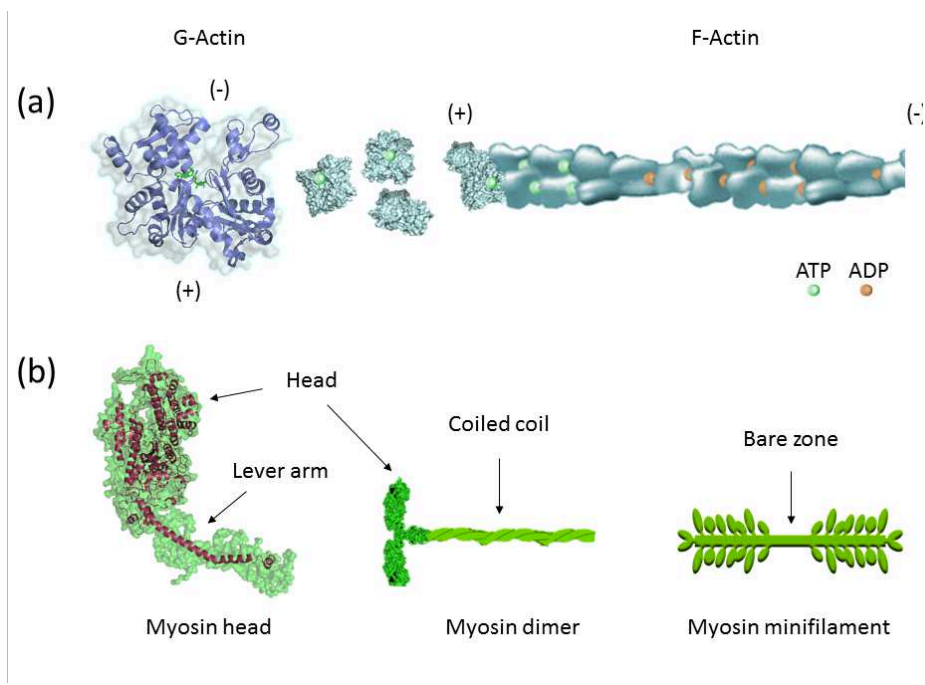}
	\caption{G-actin monomers (PDB:1J6Z) polymerize into actin filaments with a pointed $(-)$ end and a barbed $(+)$ end. Within the filament ATP is hydrolyzed to ADP. During polymerization and treadmilling monomers are added preferentially on the barbed end and released preferentially from the pointed end after hydrolyzation. (b) Myosin monomers (PDB:2MYS) are comprised of a head and tail domains. The head domain consists of an actin binding site and a lever arm, which takes part in the power stroke of the motor. The tail domains in myosin II coil to form dimers which then self-assemble into minifilaments.}
	\label{fig:actin} 
\end{figure}  

The actin monomer (G-actin) is a globular protein with a molecular mass of approximately 42 kDa\cite{Otterbein2001}. G-actin is asymmetric having a pointed end $(-)$ and a barbed end $(+)$ (see Fig.~\ref{fig:actin}(a)). In physiological conditions G-actin is found preferentially in an ATP (adenosine triphosphate) bound state. The kinetics of the actin filament growth is thermodynamically limited by the slow nucleation process; after a stable nucleus is created (with three or four monomers) filament elongation proceeds rapidly. 
The double-stranded helical actin filament retains the asymmetry of G-actin supporting a barbed and pointed end. The elongation of F-actin filaments can occur
from both ends, however the critical concentration above which growth is advantageous
is over 12 folds higher at the pointed end\cite{Wegner1983}. Starting from high monomer concentrations, the elongation process reaches a steady-state in which G-actin in ATP state is added preferentially to the barbed end and G-actin in ADP state is released from the pointed end. In cells this treadmilling process is mediated by actin binding proteins allowing for much faster turnover rates and in consequence, faster reorganization of the cytoskeleton. As a result, local control over actin binding proteins concentration allows for local changes of the turnover rate and hence to spatial heterogeneity within the cell \cite{Dominguez2011,Blanchoin2014}. 
Also in-vitro kinetics of polymerization of actin can be mediated through addition of actin binding proteins. For example, capping protein binds to the fast growing barbed end and is used, in proper stoichiometry, to control filaments length\cite{Lodish2000,Schafer1996,TDPollard2003}.

Single actin filaments are semi-flexible polymers with a persistence length of approximately $10$ \micron, a rupture force of $110$ pN, a bending energy of $4\cdot10^{21}$ J at room temperature, a buckling force of $0.4$ pN for a $1$ \micron ~long filament, and an estimated Young modulus of approximately $2$ GPa \cite{Blanchoin2014,Janmey1999}. The viscoelastic properties of an actin network are determined not only by those of the single filaments, but also by the network structure.
A semi-dilute suspension of actin filaments will form a physical gel, in which entanglements supply some resistance to shear. An addition of cross linking proteins is essential for such networks to resist flow and maintain their integrity once myosin motors are introduced\cite{Blanchoin2014,Backouche2006,Reymann2012,Haviv2008,Ideses2013}. Networks made of actin filaments support a large variety of viscoelastic properties highly dependent on the type of actin binding proteins present\cite{Lieleg2010} and on the salt concentrations in the polymerization buffer\cite{Strelnikova2016}. For example, fascin organizes actin into parallel filament bundles, while $\alpha$-actinin arranges it into antiparallel bundles. High concentrations of bundling proteins produce networks of thick bundles with much higher stiffness than networks made of crosslinked single actin filaments\cite{Lieleg2010}.

\subsubsection*{Myosin minifilaments self assembly}

One of the most important families of actin binding proteins are myosins,  the molecular motors associated with actin filaments. They are mechano-enzymes generating force by hydrolysis of ATP. This superfamily of motor proteins includes at least twenty four classes performing different roles within cells and muscles \cite{Hartman2012,Lodish2000,Syamaladevi2012}. For example, organelle transport involves myosin V \cite{Desnos2007}, muscle contraction and cytokinesis involve myosin II, which was the first to be discovered \cite{Kuhne1864} and the most studied of all myosins \cite{Lodish2000}. All Myosins are approximately $1000-2000$ residues long, and are composed of a heavy chain consisting of three domains: a conserved globular head domain (motor domain) containing an actin binding site and an ATP binding site, a flexible $\alpha$ helix neck domain associated with regulatory proteins called light chains, and a tail domain (Fig.~\ref{fig:actin}(b)). The tail domain is tailored to the function of the specific myosin, such as supporting a cargo binding site. In many myosins, including myosin II, the tail domain mediates the formation of two headed dimers by forming a long rod-like coiled coil structure binding the two monomers\cite{Hartman2012,Lodish2000,Syamaladevi2012,Reisler1980}. 
In-vitro at high salt (KCl) concentrations these dimers are stable, however at low salt concentrations multiple myosin II dimers self-assemble in an antiparallel manner to form thicker filaments termed minifilaments\cite{Reisler1980}, reminiscent of the contractile apparatus in muscle (Fig.~\ref{fig:actin}(b)) \cite{Kaminer1966,Koretz1979,Davis1988,Shutova2012}. The resulting structure of the minifilaments includes a bare zone at the center and head domains positioned at both ends of the filament. The prevailing model for minifilament self-assembly is that myosin dimers nucleate small bipolar structures\cite{Davis1982, Koretz1982, Pollard1982} that continue to grow by addition of monomers\cite{Katsura1973}. This bipolar structure is essential for myosin function in sliding two actin filaments relative to each other, and allows myosin II to function as a crosslinker between actin filaments. Control of minifilament length and number of dimers in each myosin minifilament, in vitro, is achieved by bringing the solution to a desire ionic strength. The rate at which this process is carried out may affect the assembly process. 
  \cite{Kaminer1966,Koretz1979,Ideses2013}. 
Myosin II motors use the chemical energy released in ATP hydrolysis to perform mechanical work via conformational changes in the head domain where both the ATP and actin binding sites are located. The motion is then amplified by the neck domains \cite{Hartman2012,Syamaladevi2012,Sweeney2010}.  Myosin II motors move with discrete steps of 5-15 nm reducing to 4-5 nm under load, and generating 1-9 pN forces\cite{Lodish2000,Sweeney2010,Finer1994,Ruegg2002,Kaya2013}.  Due to the multiple actin binding sites on actin filaments, a myosin minifilament can move stochastically or continuously (processively) along the actin filament. The degree of processivity of the motor motion depends on the average binding time of the myosin heads to actin (which depends on ATP concentration) and on the number of heads in each minifilament. 

\subsection{Self organization of actomyosin networks} 
\label{self}

\begin{figure}[h]
\centering
\includegraphics[scale=0.7]{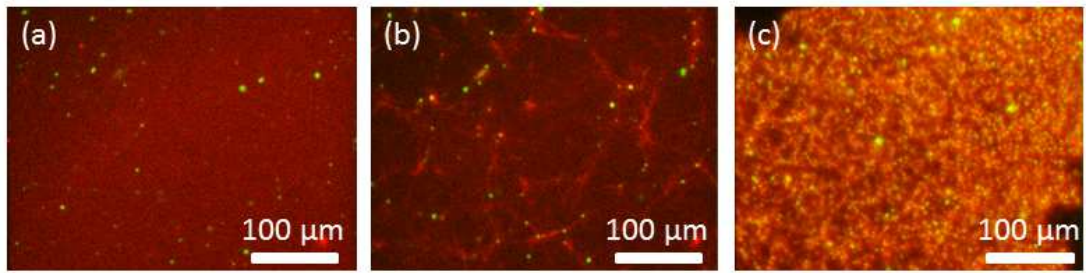}
\caption{Actomyosin network evolution. Actin is polymerized in the presence of fascin and large mysin minifilaments (150 dimers per minifilament) \cite{Ideses2013}. Fluorescent labeling allows direct observation of the actin bundles (red) and myosin minifilaments (green). (a) The network close to the initiation of polymerization, (b) after significant coarsening, (c) after 10 fold compression.}
\label{network}
\end{figure}

The minimal combination of actin monomers, myosin minifilaments and a crosslinking protein at the right salt and ATP concentration is used to create active self-organized gels in-vitro. Such polymer networks organize through a universal process of initiation, coarsening, and failure, i.e., rupture or global compression\cite{Backouche2006}. The myosin motors play an active part in this evolution process, mostly in the stages of coarsening and failure. The extent of network remodeling as well as the mode of failure may vary extensively with the concentration and relative amounts of actin, myosin, and crosslinker protein. It also depends on the type of crosslinking protein and on ATP and salt concentrations \cite{Ideses2013,Lieleg2009}. For example, in Fig.~\ref{network} the three stages of network evolution are depicted from initiation  Fig.~\ref{network}(a) through coarsening Fig.~\ref{network}(b) and failure through a 10 fold compression Fig.~\ref{network}(c).

In cells actomyosin networks appear in a large variety of morphologies well suited for their different functions. By choice of appropriate crosslinking proteins, ATP concentrations, and myosin minifilaments' concentration and size \cite{Ideses2013, Backouche2006}, many such morphologies can be obtained in-vitro. 

\subsubsection*{Initiation}
Little is known of the initial steps of network formation in in-vitro actomyosin gels due to the sub-diffraction-limit size of the actin filaments at this stage. It is known, however, that polymerization is limited by the nucleation of actin filaments, and that an addition of fascin will accelerate network formation\cite{Haviv2008a}.In the presence of fascin, myosin can further shorten the time it takes the networks to visibly form at intermediate concentrations approaching those of fascin  \cite{Ideses2013}.

\subsubsection*{Coarsening}
\label{coarsening}

Once an initial network is formed the actin network goes through a motor driven reorganization process. During this stage the thickness of actin bundles increases as does the average mesh size \cite{Bendix2008,Ideses2013}. The length scale of reorganization depends on network connectivity and on the force generation process \cite{Kohler2011}. The former depends mainly on the crosslinking protein (concentration and type), and the latter on minifilament size and the average attachment time of a single myosin head to actin. The actin architecture which is influenced by nucleation sites \cite{Reymann2012} and by crosslinker type and density affects significantly the myosin induced reorganization \cite{Smith2007}. For example, a simple system of actin and myosin can reach nematic ordering, but a trace amount of biotin/neutravidin crosslinking sites results in disordered cluster formation \cite{Smith2007}.   

The reorganization of the actin network is a result of the activity of myosin minifilaments. Myosin minifilaments anchored on two strands of actin can either slide the actin filaments apart or bring them closer. The long time result of the filament sliding and buckling is the formation of actin bundles and asters, and the pulling of excess slack in the actin network \cite{Murrell2012,Backouche2006,SoareseSilva2011}. In general, myosin activity depends both on the size of the myosin minifilaments and on ATP concentration. The processivity of the motor increases with the number of heads on a minifilament, but decreases with a decrease in the binding time of a myosin head to actin via an increase in ATP concentration. Therefore, in networks containing small minifilaments the motors can be either active, but not attached to the actin, active and attached, or attached but immobile, the probability of which depends on ATP concentration \cite{Smith2007,Vogel2013}. In addition, the total force applied by the myosin motors depends linearly on minifilament size. Finally, larger minifilaments promote larger contractile units that increase the bundling rate of the acting filaments \cite{Thoresen2013,Vogel2013}.  

The time it takes an active network to reorganize before it reaches a steady state or fails via collapse or rupture processes can range from a few minutes to several hours. During this time the network undergoes a stiffening process due to increase in actin bundles thickness and the reduction of filament bending fluctuation entropy. This gradual stiffening occurs for low and high motor concentrations as was demonstrated by using single particle microrheology \cite{Stuhrmann2012,SoareseSilva2011}. In addition, the stiffening due to internal forces applied by myosin motors was found to be equivalent to the stiffening of actin gels stretched externally \cite{Koenderink2009}.

\subsubsection*{Arrest, collapse, and rupture}
In most conditions the built up of tension due to myosin activity during the reorganization of active gels results in macroscopic compression of the gels. A recent theory identifies four competing compaction mechanisms: sarcomerelike contraction due to motors stalling at the barbed end\cite{Liverpool2005}, motion of a finite sized motor crosslinking to filaments from the intersection towards the barbed end causing  contraction\cite{Dasanayake2011}, flexible minifilaments zipping filaments together\cite{Lenz2012}, and deformable actin filaments\cite{MacKintosh2008,Mizuno2007}. This theoretical work predicts that the main mechanism for compression is the latter, namely, a local symmetry breaking in which the actin scaffold's deformation results in mesoscopic compression regardless of the sliding direction of two actin filaments generated by a myosin motor. Actin buckling due to this mechanism results from perpendicular forces rather than longitudinal buckling \cite{Lenz2014}. If the gels are held at the boundary or at high myosin concentrations, rupture occurs instead of compression. 
From monitoring the rupture process of such gels it was discovered that the final state of these gels before failure is a critically connected state \cite{Alvarado2013}. A range of network connectivity and motor activity is required to reach such a critically connected state which can then develop global compression \cite{Wang2012,Bendix2008}.

\section{Measuring stress propagation in viscoelastic materials}
\label{stress}

One way to measure the stress propagation through a given medium is to look at its mechanical response, e.g., at the displacement field resulting from a point perturbation. This concept is used, for example, in traction force microscopy\cite{Style2014}. Small tracers particles embedded in the medium have proved to be good markers for monitoring the deformation field caused by a perturbation\cite{Ladam2003}. Alternatively, stress propagations can be extracted from  the correlation in displacement of two such embedded tracer particles. When one particle is perturbed a stress field is created in the medium resulting in a displacement field entailing the other particle.  Since the second particle is moving in response to the motion of the first bead their movement is correlated. The perturbing force on the first particle may be externally or internally applied (e.g., by an external agent such as an optical or magnetic tweezers, or by a nearby molecular motor), or it can be induced by thermal fluctuations. If the beads are subjected to stochastic motion due to thermal or active fluctuations, the material's mechanical response, which is a deterministic quantity, may be distorted by the fluctuation induced noise. In such cases averaging over time and ensemble is required to characterize properly this response. 

We define ${\Delta r_\alpha (t, \tau) = r_\alpha(t+\tau)-r_\alpha(t)}$ as the vector displacement of individual tracers, where $t$ is the absolute time and $\tau$ is the lag time. The time and ensemble-averaged tensor product of the vector displacements is a measure of stress propagation, and is a function of distance and lag time:
\begin{equation}
D_{\alpha, \beta}(r, \tau) = \langle \Delta r_\alpha ^i(t, \tau)\Delta r_\beta ^j(t, \tau)\delta[r - R^{ij}(t)]\rangle_{i \not= j, t} ,
\label{dis_2p_vec}
\end{equation}
where $i$ and $j$ label different particles, $\alpha$ and $\beta$ label different coordinates, and $R^{ij}(t)$ is the distance between particles $i$ and $j$ at time $t$. Here the average is taken over the distinct terms $(i \not= j)$; the self term yields $ \langle \Delta r^2(\tau) \rangle \times \delta(r)$, the one-particle mean-squared displacement (MSD$^{\textup{1P}}$). The two-point correlation for particles in an incompressible continuum is calculated by treating each thermal particle as a point stress source and mapping its expected strain field \cite{Landau1986}. 

\begin{figure}[ht]
\centering
\includegraphics[scale=0.7]{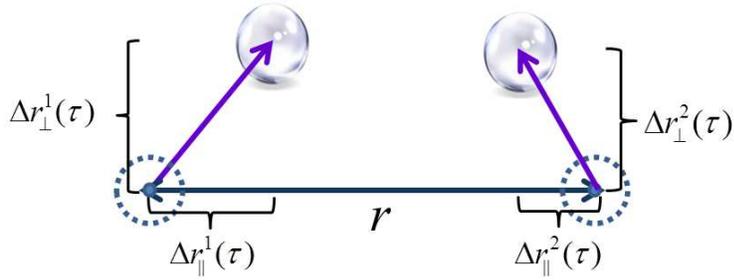}
\caption{Schematic of two-point displacement components. In this sketch, the longitudinal component $D_\parallel=\langle \Delta r^1_\parallel(\tau)\Delta r^2_\parallel(\tau) \rangle$ is the product of the displacement component projected along the line separating the tracers by distance $r$. The transverse component $D_\perp=\langle \Delta r^1_\perp(\tau)\Delta r^2_\perp(\tau) \rangle$ is the product of the displacement component projected perpendicular to the line connecting the pair. }
\label{fig:2P}
\end{figure}  

Spatially, $D_{\alpha,\beta}(r, \tau)$ can be decomposed into longitudinal $D_{\parallel}$ and transverse $D_{\perp}$ components, where the former is the component of the motion along the center-to-center separation vector of the two tracers (see Fig.~\ref{fig:2P}), while the latter is the component orthogonal to the separation vector. In an isotropic medium the off-diagonal component vanishes by symmetry. For an incompressible medium, to lowest order in $a/r$, where $a$ is the tracer particle radius, the amplitudes of the two components are related via $D_\perp=\frac{1}{2}D_\parallel.
\label{Eq:perp_para}$
Typically, $D_\parallel$ is the stronger component and hence easiest to measure in experiments from a signal-to-noise perspective.

Correlated motion measurements are used in thermodynamic equilibrium to  measure the complex shear modulus of viscoelastic media. This technique is called two-point (2P) microrheology \cite{Crocker2000}, and  was developed in 2000 as an improvement of one-point (1P) microrheology. In 1P microrheology the MSD$^\textup{1P}$ of a single tracer particle is used to extract the shear modulus of the material it is embedded in through the generalized Stokes-Einstein relation (GSER)\cite{Mason1995,Mason1997,Squires2010}. 2P microrheology takes advantage of the interparticle mechanical coupling, characterized by $D_{\alpha,\beta}(r, \tau)$ to robustly extract bulk material properties. Complex fluids contain structural elements, such as particles in a colloidal suspension or a polymer network in a gel. For these systems we may define $r_{GC}$ as a measure of the largest structural element in a complex fluid, e.g. the diameter of a colloidal particle or the mesh size of a polymer network. In a medium that is homogeneous (and isotropic) at long length-scales ($r>r_{CG}$), the strain field resulting from thermal motion of a particle is proportional to the tracer's motion and decays as $a/r$, where $r$ is the distance from the tracer.  On these scales the functional form of the decay in motion correlation is the same as in a simple incompressible fluid. This is a manifestation of momentum conservation on such scales\cite{Sonn-Segev2014}. The correlated motion of two particles with separation $r$ is driven only by modes with wavelengths greater than the separation distance. Therefore, two tracers that are separated by more than the coarse-grained length-scale $r_{CG}$ will depend on the coarse-grained, macroscopic complex modulus. At this range of separations the material is treated as homogeneous, $ D_{\parallel}(r, \tau), D_{\perp}(r, \tau)\sim r^{-1}$ within this range, and the shear modulus of the material can be determined using the relation \cite{Crocker2000}
\begin{equation}
\tilde{D}_{\parallel}(r, s) = \frac{k_BT}{2 \pi rs \tilde{G}(s)},
\label{GSER_2PM}
\end{equation}
where $ \tilde{D}_{\parallel}(r, s)$ is the temporal Laplace transform of $D_{\parallel}(r, \tau)$ and $\tilde{G}(s)$ is the temporal Laplace transform of the complex shear modulus. 

Comparing the longitudinal two-point correlation to the generalized Stokes-Einstein equation used in 1P microrheology, $\langle \Delta \tilde{r}^2 (s) \rangle = d\kT/3\pi a s \tilde{G}(s)$ in $d$ dimensions\cite{Mason1995}, suggests defining a new quantity: the two-point (2P) mean-squared displacement, MSD$^\textup{2P}$, as \cite{Crocker2000}

\begin{equation}
\textup{MSD}^\textup{2P} = \frac{2r}{a} D_{\parallel}(r, \tau).
\label{dist_msd}
\end{equation}
This is the thermal motion obtained by extrapolating the long-wavelength thermal fluctuations of the medium to the bead radius. If the material is homogeneous, isotropic on length scales significantly smaller than the tracer, incompressible, and connected to the tracers by uniform no-slip boundary conditions over their entire surfaces, the MSD$^\textup{2P}$ will match the conventional  MSD$^\textup{1P}$. Any difference between them can provide insights into the local microenvironment experienced by the tracers \cite{Valentine2001,Shin2004}. 

Here we are interested in using correlated motion to measure stress transmission between particles, i.e., the hydrodynamic interaction between them, rather than characterizing the bulk properties of the material they are embedded in. For example, the correlation in motion of two optically trapped beads suspended in water was used to measure the hydrodynamic interaction far from a boundary \cite{Meiners1999}. The hydrodynamic interaction between colloidal particles near a single rigid boundary was calculated \cite{Blake1971,Pozrikidis1992} and measured\cite{Dufresne2000} for a pair of particles diffusing at a distance $h$ above a wall  :
\begin{eqnarray}
D_{\parallel}(r \gg h) &=&  \frac{3\kT h^2}{2\pi \eta r^3 },\\ D_{\perp}(r \gg h) &=& \frac{3 \kT h^4}{ 4\pi \eta r^5}.
\label{Eq:cordif_wall}
\end{eqnarray}   
These coefficients describe the leading terms of the in-plane correlated diffusion between two colloidal particles. Note that the leading term in the hydrodynamic interaction decays as $\sim r^{-3}$, rather than $\sim r^{-1}$, which is due to the unconserved momentum in the system. Similarly, in other confined geometries hydrodynamic interactions depend differently on distance, yielding different functional forms for the stress transmission \cite{Diamant09}, e.g. $ D(r, \tau) \sim~r^{-2}$ in quasi-two dimensional samples \cite{Cui2004}. This was demonstrated experimentally for different colloidal suspensions \cite{Cui2002,Cui2004,Sonn-Segev2015}.  

In order to extract the full information hidden in the correlated motion of tracer particles it is beneficial to compare measurements to a physical model describing the embedding material. For example, the thickness of a soap film\cite{Prasad2009} or a thin viscoelastic layer\cite{Ladam2003} could be extracted given a proper model for deformations in a quasi 2D layer with free or rigid boundaries, respectively.   

\section{Stress propagation in passive in-vitro actomyosin networks}
\label{passive}
\subsection{The measurements: 2P correlations}

\begin{figure}[h!]
  \includegraphics[scale=0.55]{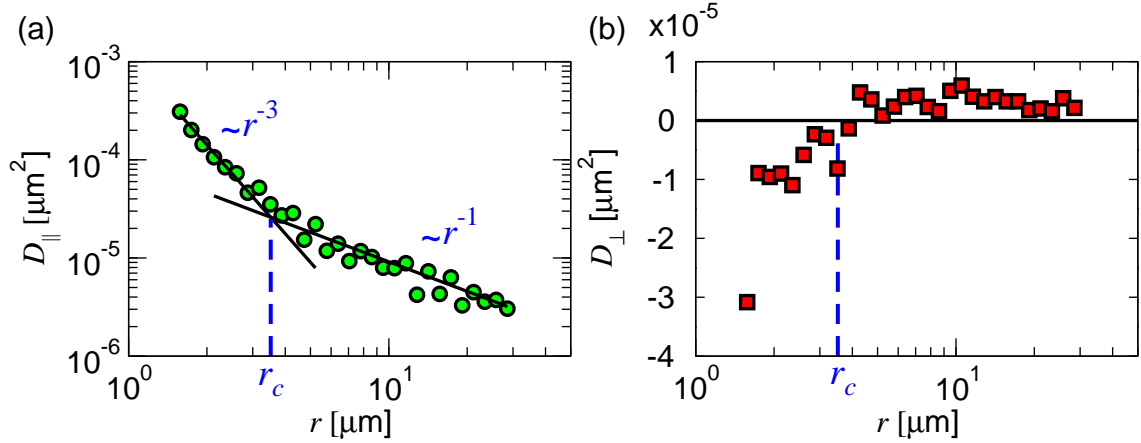}
  \caption{(a) Longitudinal and (b) transverse displacement correlations as a function of particle separation, $r$, at lag time $\tau=0.014$s for $\xi_s=0.3~\mu$m,  $a=0.245~\mu$m and $\la =13~\mu$m. The crossover distance $r_c$ (blue dashed line) is defined at the intersection of the fitted dominant ($r^{-1}$) and subdominant ($r^{-3}$) power-law decays of $D_\parallel$. Reproduced from \cite{Sonn-Segev2014a} with permission from The Royal Society of Chemistry.}
  \label{fig:Drq}
  \end{figure}

Let us start by considering stress propagation in an entangled actin networks with a mesh size $\xi_s=300$ nm \cite{Sonn-Segev2014,Sonn-Segev2014a}. In  Fig.~\ref{fig:Drq} the correlated motion in the longitudinal and transverse direction, $D_{\parallel}$ and $D_\perp$, respectively,  are presented as a function of particle separation. There are several interesting features in these plots: (i) there are two regimes of stress propagation as a function of inter-particle separation, (ii) the crossover distance between the two regimes, $r_c$, is an order of magnitude bigger than the particle diameter (0.49 \micron) and the mesh size, (iii) the new intermediate regime is characterized by $D_\parallel\sim r^{-3}$ and $D_\perp<0$, as opposed to the well known long distance scaling i.e. $D_{\parallel,\perp}\sim r^{-1}$.

These results can be interpreted by reexamining the Stokes problem of a rigid sphere of radius $a$ driven by a constant force $\vec{F}$ through an incompressible fluid of viscosity $\eta$ \cite{happel91}. The fluid velocity at a distance $r$ from the sphere can be described by a multipole expansion of the force and density fields, in analogy to the multipole expansion commonly done to describe the electrical field arising from a charged sphere. The first term is a force monopole, which is the field that would arise from the perturbed sphere (colloid) if it was infinitely small. The second contribution would have been a force dipole, but for this scenario i.e. a sphere in an isotropic medium, this term vanishes \cite{Pozrikidis1992}. The third term in the force field is a force quadrupole; its physical meaning is that there is a difference between the force field created by a point particle and one with finite size. The first term in the mass field is a dipole; due to conservation of mass a local increase in density must be combined with a decrease in density nearby. As opposed to the force monopole, which decays as $r^{-1}$, the two subdominant terms in the flow field, the mass dipole and force quadrupole decay as $r^{-3}$ with different signs \cite{Sonn-Segev2014,Diamant2015}. Although the functional form of these two contributions is the same, their physical origin is different, as we discuss below. In a simple fluid, such as a Newtonian fluid, the subdominant response becomes significant only at distances comparable to the particle size $a$, the only length scale in the system. Therefore, the subdominant response decays as $a^2/r^3$, and vanishes as $a\rightarrow0$. For a viscoelastic complex fluid the two subdominant contributions become separate,  the mass dipole, originating from mass conservation and reflecting the fluid flow in the vicinity of the particle, depends on the local environment of the tracer, while the force quadrapole reflecting momentum transfer through the bulk material depends on the bulk viscosity\cite{Sonn-Segev2014,Diamant2015}. In a case where the local environment (solvent) has much lower viscosity than the bulk material viscosity (polymer network), we expect the subdominant contribution to manifest to much larger distances. 

For example, even in the case of entangled actin gels (Fig.~\ref{fig:Drq}) with no addition of cross-linkers or bundling agents we observe the subdominant response up to a distance of approximately $r_c=3.5$ $\mu$m. The prediction in such a case is that:
\begin{eqnarray}
D_\parallel \sim r^{-3} \textup{  and  } D_\perp<0 \textup{  if  } r<r_c \nonumber\\
D_\parallel \sim r^{-1} \textup{  and  } D_\perp>0 \textup{  if  } r>r_c.
\end{eqnarray}
This result should hold for any complex fluid with $r_c \sim \eta_b/\eta_\ell$, where $\eta_b$ ($\eta_\ell$) corresponds to the bulk (local) viscosity. Note that for complex fluids, with more than one relevant length scale (e.g. $a,\xi$ for a polymer gel), the response decays as $~\xi^2/r^3$ and therefore does not vanishes as $a\rightarrow0$. In this context we call the effective viscosity experienced by the tracer particle due to complex media (for example, solvent and the polymer network) the local viscosity. The local viscosity experienced by a tracer particle in an actin network decreases with particle size becoming closer to the solvent viscosity (although it will never reach exactly that limit).  

In Fig.~\ref{fig:Drq} the theoretically expected power law is seen only for half a decade. Nonetheless, we have observed this exact power law at different lag times in the same experiments, and in numerous other actomyosin networks: with and without cross-linking molecules, with and without myosin, and at different filament lengths. Further support for this power law dependence is that it can be derived base on the condition of mass conservation \cite{Diamant2015}. Finally, the theoretical analysis discussed above is confirmed below (Fig.~\ref{fig:figure4}), where a rescaling of the data according results in a single master curve.

We note here that the asymptotic behavior  of the correlated motion of beads in actin networks has been measured previously, and used to demonstrate the advantage of 2P microrheology in measuring the bulk shear modulus of viscoelastic inhomogeneous materials \cite{Crocker2000,Gardel2003}. It was also shown that in such complex fluids intrinsic structural length scales affect the materials' shear modulus \cite{Liu2006}. 


\subsection{The interpretation: the two fluid model}
In order to understand the mechanical response of actin gels in the intermediate regime we require a theoretical model for such gels. We use the two fluid model of polymer gels \cite{DeGennes76,DeGennes76a,Milner1993,Levine2000,Levine2001} for this purpose. In this model the polymer is treated as a dilute viscoelastic network coupled to an incompressible solvent by friction forces. A local mechanical perturbation by a tracer particle will cause the solvent to flow through the polymer network in its proximity. However, at some larger distance, friction forces will cause the polymer network to move together with the fluid, as one continuum medium. There will arise a typical distance separating the flow of fluid against and with the polymer network. This crossover distance, $r_c$, can be calculated within the framework of the two fluid model \cite{Sonn-Segev2014,Sonn-Segev2014a,Diamant2015}, and reads:

\begin{equation}
  r_c = a [2(\eta_b/\eta_\ell)g(\xi_d/a)]^{1/2},\ \
  g(x) = x^2+x+1/3,
\label{rc}
\end{equation} 
where $\xi_d$ is the dynamic correlation length of the viscoelastic gel.

\begin{figure}[h!]
\centering
\includegraphics[scale=0.65]{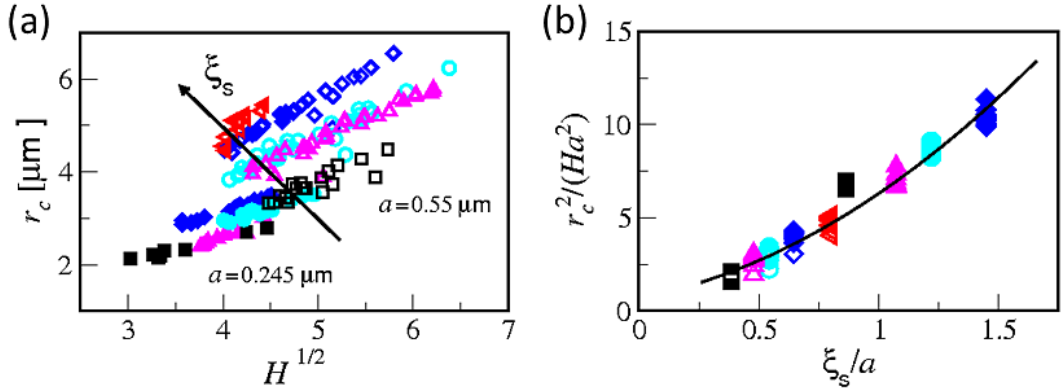}
\caption{Crossover distance for experiments on passive entangled actin networks at different conditions. (a) For all
  conditions $r_c$ is linear with $\sqrt{H}$ and increases with either
  $\xi_s$ or $a$. (b) All experimental results fall on a master curve
  once $r_c^2$ is normalized by $Ha^2$ and rescaled according to Eq.~(\ref{rc}). A fit to Eq.~(\ref{rc}) with $\xi_d=1.25\xi_s$ is presented by the solid line. Open (filled) symbols correspond to $a=0.55$ ($0.245$) $\mu$m. Each symbol corresponds to a different mesh size: $\xi_s=0.21$ (black squares), $0.26$ (magenta triangles), $0.3$ (cyan circles), $0.35$ (blue diamonds), and $0.44$ $\mu$m (red left triangles). The average filament length for all experiments was $\la=13~\mu$m. Adapted from \cite{Sonn-Segev2014} with permission from the American Physical Society.}
\label{fig:figure4}
\end{figure}

The ratio between the bulk and local viscosity is equivalent to the experimentally measured ratio $H(\tau)=\textup{MSD}^\textup{1P}/\textup{MSD}^\textup{2P}$ which is time (and frequency) dependent in a viscoelastic material \cite{Sonn-Segev2014,Sonn-Segev2014a}. This is true, in thermal equilibrium, since $\textup{MSD}^\textup{1P}$ is inversely related to the local viscosity of the network, and $\textup{MSD}^\textup{2P}$ reflects bulk properties\cite{Sonn-Segev2014a}. Therefore, $r_c \sim H(\tau)^{1/2}$, as was demonstrated in passive actin networks with various particle sizes and mesh sizes (Fig.~\ref{fig:figure4}(a)). 

In actin $H\sim 100$ \cite{Sonn-Segev2014a}, causing $r_c$ to be ten times larger than the typical length scales in the system ($\xi_s,a$). This means that the mechanical response of actin networks decays faster than originally expected crossing over to a slower decay rate at a distance of a few micrometers (In Fig.~\ref{fig:figure4} $r_c$ ranges between 2 \micron to 6.5 \micron). This decay length is comparable to the size of biological cells. For stiffer cytoskeleton networks containing microtubules $r_c$ can reach values of 15 \micron \cite{Pelletier2009}.

We can take our analysis one step further and extract structural information from the stress transmission signal, since the intermediate term includes the information about the correlation length of the network. (Eq.~\ref{rc}). The missing piece of information in Eq.~\ref{rc} is the ratio between the dynamic correlation length and the structural length scale, i.e., the mesh size. The dynamic correlation length is defined as the length scale over which momentum is absorbed in the system. It is related to the gel's mesh size, but is not equal to it. This was obtain from the fit to the data in Fig~\ref{fig:figure4}(b) to be, $\xi_d/\xi_s=1.25$. Using this relation we can now extract the mesh size of an actin network from $r_c$. As stated in Sec.~\ref{stress}, the combination of stress propagation measurements and its modeling reveals structural information on the sample.

\begin{figure}[h!]
\centering
\includegraphics[scale=0.4]{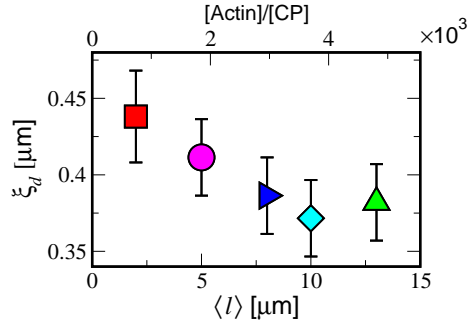}
\caption{Dynamic correlation length, $\xi_d$, as a function of the average filament length, $\la$ (bottom) and actin/CP concentration ratio (top). Actin concentration was held at 1 mg/ml, resulting in a $\xi_s=0.3~\mu$m, and $a=0.245~\mu$m. Reproduced from \cite{Sonn-Segev2014a} with permission from The Royal Society of Chemistry }
\label{fig7}
\end{figure}

It can be argued that many complex fluids and especially cytoskeleton networks possess more than one typical structural length scale. For example, actin filaments have a persistence length of $5-10$ $\mu$m and a contour length that can vary between 2 $\mu$m to $20$ $\mu$m. The effect of filament length was studied recently \cite{Sonn-Segev2014a}. Following the analysis described above, the stress propagation signal ($D_\parallel$) and the viscosity ratio $H(\tau)$ were measured for networks made of filaments of well defined length (ranging between 2 to 13 \micron), all with a mesh size of $\xi_s=0.3$ \micron. The dynamic correlation length $\xi_d$ was extracted by scaling $r_c$ with $H$ and $a$ for all the networks.  Fig~\ref{fig7} shows that for filaments shorter than $8$ $\mu$m,  $\xi_d$ depends also on filament length for a given $\xi_s$, i.e, stress propagation depends on also on $xi_d$.   

\section{Stress propagation in steady state in vitro actomyosin networks}
\label{active}

The study of stress propagation by 2P microrheology requires large amounts of data. Such measurements are ideally done in thermodynamic equilibrium and are challenging in rapidly evolving networks. To study of stress propagation in active mater, such as actomyosin networks it is advantageous to work in conditions where the gel arrives at a long-lived active steady state.
  
\subsection{Formation of active steady state networks} 

Actomyosin networks which arrive at long lived active steady states were created recently by polymerizing a mixture of unlabeled and biotinilated actin monomers in the presence of neutravidin and small myosin minifilaments. An average distance between crosslinkers of $ \approx 3~\mu$m, and a mesh size of $\xi_s=0.3~\mu$m  were obtained by stochiometry. Varying degrees of activity were achieved by changing myosin minifilament concentration and size ($N_{myo}=19\pm3$ or $N_{myo}=32\pm5$  two-headed myosin dimers per minifilament) \cite{Future}. The number of myosin heads per minifilament was estimated from the distribution of minifilament length as measured using CryoEM (see supplementary information in \cite{Ideses2013}).

\begin{figure}[h]
	\centering
	\includegraphics[scale=0.32]{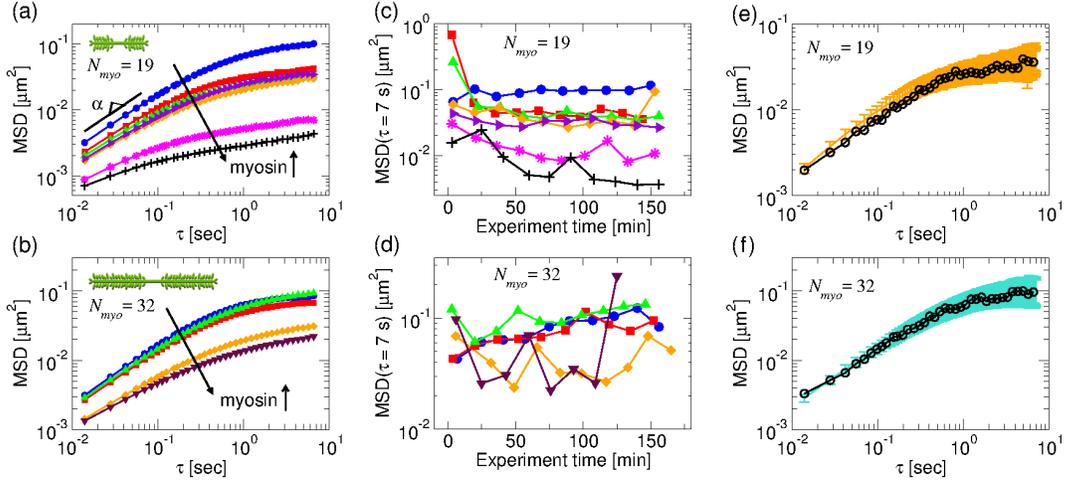}
	\caption{Mean squared displacement (MSD) of particles in networks with different [myosin]/[actin] at two minifilaments sizes. (a) and (b) time and ensemble-averaged MSD of probe particles as a function of lag-time $\tau$ approximately 100 mins after polymerization. Minifilaments are composed of $N_{myo}=19$ (a) or $N_{myo}=32$ (b) myosins heads. (c) and (d) MSD at a lag time of $\tau=7~$s re-measured as a function of experiment time. The experiment time is the time between the onset of gel polymerization and the measurement time. Sizes of mini-filaments are $N_{myo}=19$ (c) and $N_{myo}=32$ (d). Colors and symbols correspond to different [Myosin]/[Actin] ratios: 0 (blue circles), 0.0017 (red squares), 0.0025 (green triangles), 0.005 (orange diamonds), 0.0083 (violet right triangles), 0.01 (maroon down triangles) 0.012 (magenta stars) and 0.02 (black pluses). (e) and (f) Comparison between time-averaged and ensemble-averaged MSD for networks with [Myosin]/[Actin]=0.0025 approximately 100 min after polymerization.  Sizes of mini-filaments are $N_{myo}=19$ (e) and $N_{myo}=32$ (f), and initial slopes are $\alpha=0.7\pm 0.05$ and $\alpha=0.8\pm 0.05$. }
	\label{fig:characteristics}
\end{figure}

In order to determine when these gels reach a steady state the motion of fluorescent polystyrene beads ($a= 0.55$ \micron) was recorded for several hours at intervals of 15~min. The most obvious effect of myosin concentration on these actomyosin networks is to increase their stiffness, as seen from the decrease in the  MSD$^{\textup{1P}}$  of the tracer particles with the increase in myosin concentration (Fig. ~\ref{fig:characteristics}a,b). 
About 50 min after mixing the various components the MSD$^{\textup{1P}}(\tau=7 ֿֿ \textup{s})$ settled to a steady value for most of the myosin concentrations (Fig. ~\ref{fig:characteristics}c,d).  
The ensemble and time average of the MSD$^{\textup{1P}}$ were compared to demonstrate that although the system is not at thermal equilibrium it is ergodic (Fig. ~\ref{fig:characteristics}e,f) \cite{Future}.

\subsection{Effect of motor concentration and minifilament size} 

\begin{figure}[h]
	\centering
	\includegraphics[scale=0.45]{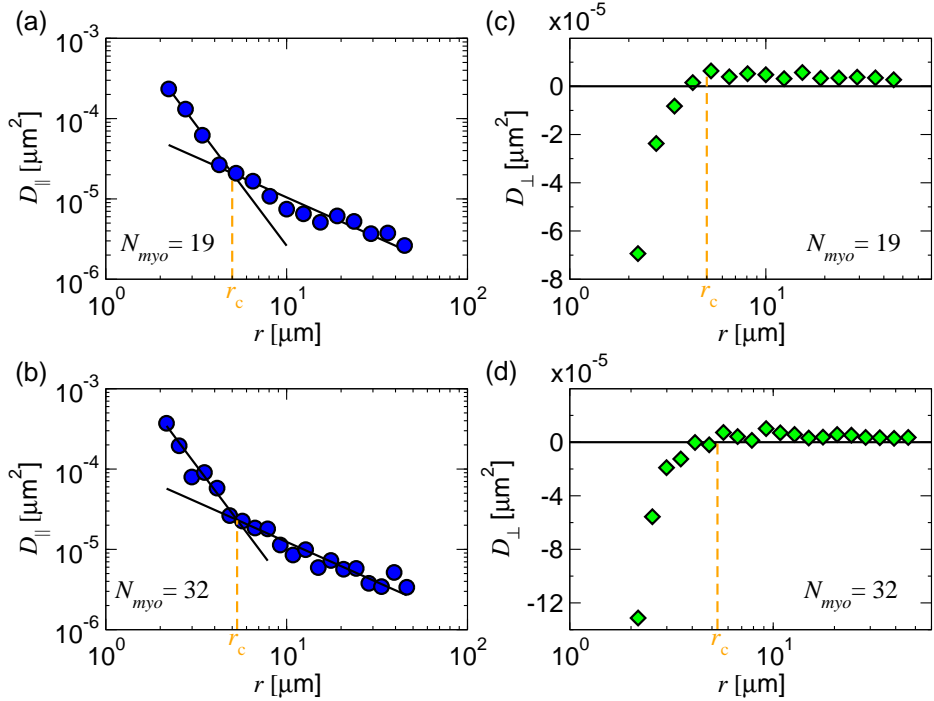}
	\caption{  Correlated motion of beads in actomyosin active networks. (a),(b) Longitudinal displacement correlations as a function of particle separation at lag time $\tau=0.014$~s and [Myosin]/[Actin]=0.0025.  Mini-filaments are constructed by $N_{myo}=19$ (a) or $N_{myo}=32$ (b) myosins heads. The cross-over distance (orange dashed line) is clearly seen as in passive actin networks.   (c) and (d): Transverse displacement correlation at the same conditions as in (a) and (b). }
	\label{fig:rcross}
\end{figure}

The two fluid model description, that was used for the passive networks, should hold for this system as well. Here, active random fluctuations are present in addition to the thermal fluctuations, but the system is still comprised of a polymer network immersed in a solvent. The significant amount of stiffening in these networks due to the motor concentration (Fig.~\ref{fig:characteristics}a,b) can be attributed to the addition of actin crosslinking sites by myosin minifilaments and to the reduction of slack in the actin filaments \cite{Future}. However, these changes do not affect the functional form of stress propagation through the active networks (Fig.~\ref{fig:rcross}). As in passive actin networks (Fig.~\ref{fig:Drq}), the correlated diffusion in the longitudinal direction decays fast at short distances $D_\parallel\sim r^{-3}$, and slowly at large distances $D_\parallel\sim r^{-1}$ .

\begin{figure}[h]
	\centering
	\includegraphics[scale=0.45]{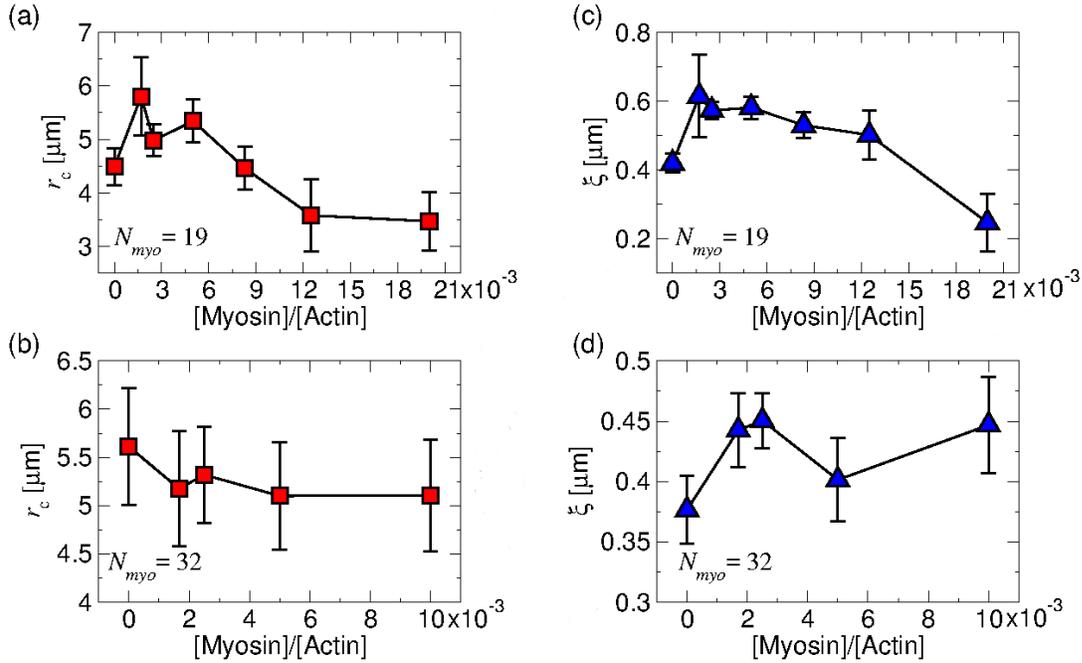}
	\caption{(a) and (b) The cross-over distance, $r_c$, in networks with increasing myosin concentration. (c),(d) The Dynamic correlation length  $\xi_d$ as a function of myosin concentration, base on the two fluid model scaling (Eq.~\ref{rc}). }
	\label{fig:rcross1}
\end{figure}

The crossover distance, $r_c$ for both myosin minifilament sizes changes slightly with myosin concentration (Fig.~\ref{fig:rcross1}).  It ranges  between $4.5-3.5$ $\mu$m  and $5.5-5.0$ $\mu$m for $N_{myo}=19$ and $32$ respectively (Fig.~\ref{fig:rcross1}).  $r_c$ for the large minifilaments is bigger than for the smaller minifilaments. Considering that a polymer becomes stiffer with applied stress \cite{Rubinstein2003} due to a reduction in its configuration entropy, this result is expected, since larger myosin minifilaments can apply stronger forces on the network making it much stiffer and as a consequence increasing $\eta_b/\eta_\ell$. Moreover, at the same  myosin to actin concentration ratio, which are kept constant in the experiments, they add less crosslinking sites resulting in a larger mesh size.

The structural features of these active networks were smaller than the diffraction limit and could not be resolved by optical microscopy, since no bundling proteins were added during preparation. Therefore, a direct observation of the structural evolution of these networks, which is expected for actomyosin networks (see Sec.~\ref{coarsening}), was not possible here. Nonetheless, insight into the structural evolution of the networks after initiation of polymerization and before they reach steady state can be obtained from $r_c$, based on the two fluid model,  Eq.~\ref{rc}, and the assumption that the networks are close enough to thermal equilibrium for the extracted $\xi_d$ to be a good estimate of $\xi_s$. For both systems we see a jump in the dynamic correlation length with addition of myosin minifilaments. We believe this result reflects the expected coarsening of the network due to the presence of motors prior to arriving at a steady state. It was recently shown that myosin stiffens actin networks in two ways, one of which is the by addition of crosslinking sites to the network \cite{Future}. As a result the mesh size and dynamical correlation length are expected to decrease. This effect can be seen in the $N_{myo}=19$ system for myosin concentrations [myosin]/[Actin]$>6\cdot 10^{-3}$. Here, the myosin minifilaments concentration becomes comparable to the concentration of biotin/neutravidin crosslinking sites. 

\section{Stress propagation in evolving in-vitro actomyosin networks}
\label{evolving}

\begin{figure}[h]
\centering
\includegraphics[scale=0.55]{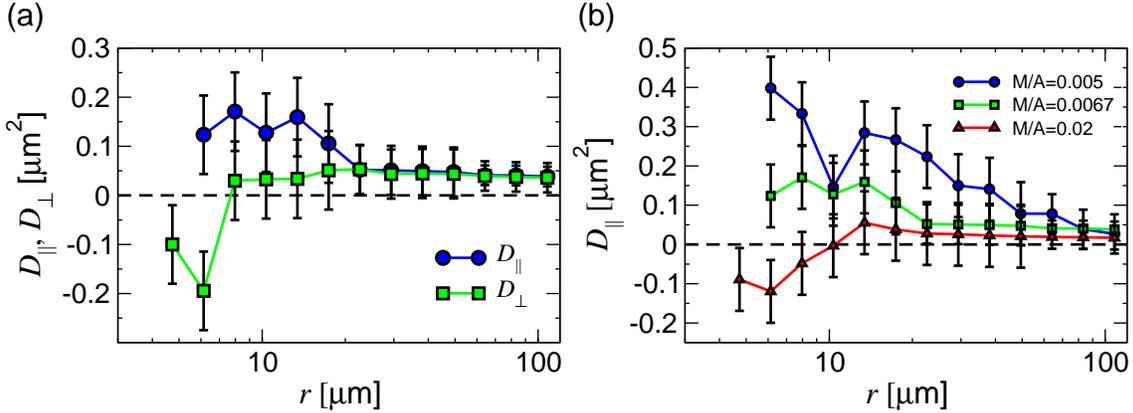}
\caption{Correlated motion of myosin minifilaments embedded in the actin network (as in \cite{Ideses2013}) (a) $D_{\parallel}$ and $D_{\perp}$ of an evolving network  280 s after initiation with [myosin]/[actin]$=0.0067$. (b) $D_{\parallel}$ at various myosin concentrations (M/A denotes myosin to actin concentration ratio).}
\label{fig:Dii_ss_1}
\end{figure}

So far we have considered stress propagation through the bulk of a complex fluid. It is also interesting to ask how stress propagates directly through the polymer network (and not through the solvent). It is known that the cytoskeleton (i.e. actin filaments) is connected to the extracellular matrix through binding sites\cite{Cooper2000}. A mechanical signal, passing from the extracellular matrix through these adhesion points into the cell may propagate along the actin network as well as through the bulk. To characterize stress propagation through the actin network we turn to study evolving actomyosin networks  comprised of actin,  fascin, and large myosin minifilaments ($N_{myo}=150$) \cite{Ideses2013}. In this actomyosin system (see Fig.~\ref{network}(a)) the fluorescently labeled actin (red) and myosin (green) can be directly observed. These gels follow the evolution stages described in Sec.~\ref{self} and Fig.~\ref{network}. After approximately two minutes these gels are fully connected and start coarsening. During the whole coarsening stage the myosin minifilaments remain embedded in the actin network. The correlated motion ($D_{\parallel}, D_{\perp}$) of myosin minifilaments, averaged over the entire coarsening stage is presented in Fig.~\ref{fig:Dii_ss_1}(a). As opposed to the measurements of stress propagation between beads which are not attached to the network, there are very strong correlations in the motion of the myosin minifilament. Thus, a relatively low amount of measurements is required to extract reproducible results and good signal to noise ratio, at least for ubiquitous particle separations of the order of ten of micrometers. 

The functional dependence of $D_{\parallel}$ and $D_{\perp}$ on distance is very different from the bulk response. Connection through and elastic network results in positive motion correlation up to large distances both in the longitudinal and transverse directions (Fig.~\ref{fig:Dii_ss_1}(a)). This long range correlation reflects the motion of the network's center of mass and should therefore persist throughout the sample. Statistics are insufficient to obtain a reliable measure of the functional form of this long range response. At small inter-motor separations statistics is also low, since we observe, most commonly, separations of 5-50 $\mu$m. Interestingly, in cases where the short distance response measurement is reliable, i.e., $D_{\perp}$ in Fig.~~\ref{fig:Dii_ss_1}(a) and $D_{\parallel}$ of M/A=0.02 in Fig.~~\ref{fig:Dii_ss_1}(b), $D_{\parallel}$ and $D_{\perp}$ increase at short distances to some maximal value before decreasing down to the long range value.  At very short distances negative correlations may arise. These can be attributed to a lack of statistics, to crosslinking sites that sustain local torques, or to local contraction effects within the network between nearby motors.

We focus on the stronger signal $D_{\parallel}$ to demonstrate the effect of motor concentration (Fig.~\ref{fig:Dii_ss_1}(b)). A similar behavior of $D_{\parallel}$ as a function of distance is seen for the various myosin concentrations, and the correlations at large distances are essentially equal. However, at intermediate distances, the lower the myosin concentration the higher the 2P correlations. This result probably reflects the higher probability of the two motors to be affected by the same third motor at smaller motor concentrations, as the inter-motor distance increases with the decrease in motor concentration. We estimate the inter-motor distance $\Delta x_m\approx 10,30,40~\mu$m for [myosin]/[actin]=0.02,0.0067,0.005 respectively, which supports this interpretation.

\section{Conclusions}

In this review we suggest two point motion correlation of embedded beads as a measure of stress propagation through complex materials. We demonstrate our approach in studying stress propagation in model cytoskeleton networks including actin gels at thermal equilibrium, actomyosin networks which arrive at a mechanical steady state, and fast evolving actomyosin networks. Furthermore, we show that structural information is encoded into the stress propagation signal and can be  extract from experiments by comparison to the solution of the stokes problem in a relevant model for the specific complex fluid in question.

Stress propagation was characterized here both through the bulk material (Sec.~\ref{passive} and \ref{active}), and directly through the polymer network (Sec.~\ref{evolving}). The amount of statistics required in order to get a good signal of stress propagation through the polymer network was several orders of magnitude smaller than what was required to recover the propagation signal through the bulk. This is due to the much stronger correlations in motion of two tracers connected directly to an elastic object. In cells, where mechanical signals are commonly used, it is convenient that perturbations applied on the cytoskeleton protein directly propagate well to long distances, while their effect on the surrounding fluid decays fast ($\sim r^{-3}$). Long range perturbations generated by myosin II motors were previously proposed to promote dynamic motor-mediated attraction and fusion of actin bundles. These perturbation were suggested to propagate via a 2D elastic actin network to which the bundles are coupled and via the surrounding fluid\cite{Gilo2009}.

The work reviewed here focused on the linear response of a material to thermal and small active perturbations. Cytoskeleton networks, however, have non-linear elastic properties \cite{Gardel2006}, as do many complex fluids (e.g., \cite{Waitukaitis2012}). A natural extension of our approach is to actively and strongly perturb one of the tracer particles to characterize the non-linear response of complex fluids, as done in active microrheology. In this method a tracer particle is externally driven, for example by means of a magnetic field \cite{Amblard1996,Choi2011} or optical tweezer \cite{Mizuno2007,Mizuno2008,Lee2010}. The response of the media is then separated to an in-phase part reflecting the elastic shear modulus and an out-of-phase part which reflect the loss modulus. The control over the amplitude and strain rate of the mechanical perturbation is essential for probing the non-linear response of the material.  There are a few reposts on active microrheology in actin networks. It was shown that active microrheolgy of actin networks at low strain amplitude agrees well with passive microrheology measurements \cite{Mizuno2008,Lee2010}. As expected, at larger strain amplitudes, inaccessible to passive microrheology, the known non-linear stiffening of actin networks \cite{Gardel2006} can be observed by active microrheology \cite{Lee2010}. Active microrheology of active actomyosin networks was used to characterize the athermal fluctuations of such networks \cite{Mizuno2007} at the linear mechanical response regime. Generalizing our approach to further study actin networks focusing on their non linear response with and without myosin is expected to yield a better understanding the nature of stress transmission in these systems.

\section*{References}


\providecommand{\newblock}{}

\end{document}